# Optical properties of silicon-implanted polycrystalline diamond membranes

H. Kambalathmana,[a,b] A. M. Flatae,[a,b] L. Hunold,[a,b] F. Sledz,[a,b] J. Müller,[b,c] M. Hepp,[b,c] P. Schmuki,[d] M. S. Killian,[b,e] S. Lagomarsino,[a,b,f] N. Gelli,[f] S. Sciortino,[f,g] L. Giuntini,[f,g] E. Wörner,[h] C. Wild,[h] B. Butz,[b,c] and M. Agio[a,b,i,*]

[a] Laboratory of Nano-Optics, University of Siegen, 57072 Siegen, Germany
[b] Center for Micro- and Nanochemistry and Engineering (Cμ), University of Siegen, 57068 Siegen, Germany
[c] Micro- and Nanoanalytics Group, University of Siegen, 57076 Siegen, Germany
[d] Chair for Surface Science and Corrosion, FAU Erlangen, 91058 Erlangen, Germany
[e] Chemistry and Structure of novel Materials, University of Siegen, 57076 Siegen, Germany
[f] Istituto Nazionale di Fisica Nucleare, Sezione di Firenze, 50019 Sesto Fiorentino, Italy
[g] Department of Physics and Astronomy, University of Florence, 50019 Sesto Fiorentino, Italy
[h] Diamond Materials GmbH, 79108 Freiburg, Germany
[i] National Institute of Optics (INO), National Research Council (CNR), 50125 Florence, Italy

**Abstract:** We investigate the optical properties of polycrystalline diamond membranes containing silicon-vacancy (SiV) color centers in combination with other nano-analytical techniques. We analyze the correlation between the Raman signal, the SiV emission, and the background luminescence in the crystalline grains and in the grain boundaries, identifying conditions for the addressability of single SiV centers. Moreover, we perform a scanning transmission electron microscopy (STEM) analysis, which associates the microscopic structure of the membranes and the evolution of the diamond crystals along the growth direction with the photoluminescence properties, as well as a time-of-flight secondary ion mass spectrometry (ToF-SIMS) to address the distribution of Si in implanted and un-implanted membranes. The results of the STEM and ToF-SIMS studies are consistent with the outcome of the optical measurements and provide useful insight into the preparation of polycrystalline samples for quantum nano-optics.

---

[*] Corresponding author. Tel: +49 271 740 3532. E-mail: Mario.Agio@uni-siegen.de (Mario Agio)

**1. Introduction**

Solid-state quantum emitters are promising single-photon sources for novel applications including quantum information and sensing (1–6). Among them, the negatively-charged silicon-vacancy (SiV) color center in diamond (1,7–11) is a unique resource due to its optical properties, such as a strong emission in the zero-phonon line (ZPL), photon indistinguishability, high repetition rate, the possibility to operate at room and higher temperatures, and its controlled creation. However, the poor out-coupling efficiency of SiV centers due to the high refractive index of diamond and the position of the color centers in the host may hinder the efficient optical interaction needed for emerging quantum science applications.

Nano-photonic structures, such as solid-immersion lenses (12), nanopillars (13,14), and photonic-crystal cavities (15–17) have shown remarkable improvements in the coupling efficiency and/or the emission rate, but they require advanced nanofabrication tools. Alternatively, color centers in thin diamond membranes would more easily allow to manipulate their radiation pattern for bright directional emission (18,19), the realization of ultrafast single-photon emission (20), and also, similarly to two-dimensional materials (21), enable hybrid integration into nanophotonics devices, which so far has been mostly pursued using diamond nanocrystals (22,23).

While single-crystal diamond membranes require sophisticated technological approaches and so far did not achieve thicknesses below 100 nm (24,25), fabricating thin polycrystalline diamond (PCD) membranes represents a viable approach. Over the past 20 years, a variety of diamond synthesis techniques have been developed. Among them, microwave plasma chemical vapor deposition (MPCVD) has emerged as the preferred fabrication method for high-quality thin diamond membranes, with control over the size and geometry (26,27).

Owing to the growth process, the grain structure of PCD membranes has a non-uniform composition compared to other polycrystalline structures. The crystal quality and the texture of the diamond film are tangled with a few key parameters, such as the material of the substrate used, the substrate temperature, and the ratio of process gas mixture (28). Generally, in an MPCVD process, decreasing the $CH_4$ content in the mixture of $CH_4/H_2$ gas leads to PCD films with micrometer size grains, while, increasing the $CH_4$ content leads to diamond films with nanocrystalline grains (29). Accordingly, the morphology of the crystal is modified from faceted microcrystals to ball-shaped nanocrystals. Moreover, the ratio of the $CH_4/H_2$ gas mixture defines the quality of CVD diamond film by balancing the $sp^3/sp^2$ content in the crystal (30,31). Furthermore, choosing an appropriate substrate for crystal growth is imperative. The MPCVD process on a silicon substrate initially forms a carbide interfacial layer (which reduces the stress at the interface) followed by the desired diamond growth. The growth process generally takes place at elevated temperatures, but, upon cooling, the compressive stress resulting from the thermal expansion coefficient mismatch between the substrate and the diamond can induce structural damages to the film (28).Generally, silicon atoms are already present in most of the CVD fabricated diamond films via diffusion from a silicon-containing substrate and/or from the silica reactor windows (32). This process creates SiV centers without further annealing and ion implantation procedures (33). By balancing a dopant source, the depth of impurities can be controlled in this method. Sometimes it is also possible to ensure that up to one SiV center appears per grain provided the growth rate is known. However, this method does not have lateral control of the SiV creation and leads to clustering formation of SiV centers (34).

Using a focused ion beam it is possible to implant the ions in the desired position with fine control over the number of ions implanted (9,11,35). However, it is necessary to restore the diamond lattice and to activate the SiV centers after implantation by promoting both vacancy-interstitial and silicon-vacancy recombination via thermal activation of vacancy diffusion in a high vacuum chamber (thermal annealing). The controlled positioning of the vacancy centers on the sample is the leverage of this technique, however at the expense of random activation and orientation of the centers.

At first, we investigated the optical properties of SiV centers in PCD membranes of thicknesses from 3 μm down to 55 nm fabricated using MPCVD on a silicon wafer (20). However, isolating single emitters was limited by the dimensions of the single grains and by the fluorescence background originating from the grain boundaries (GBs). On the other hand, the deposition of thicker PCD membranes would create larger grains, thereby the emitter could be located within the grain centers (GCs) with a higher signal-to-noise ratio suitable for addressing single color centers. It is therefore critical to understand the relationship between the morphology of μm-thick PCD membranes and the optical properties of SiV centers.

In this work, we investigate 5-μm thick PCD membranes to gain insight on the possibility to use them for experiments and applications based on single color centers, especially the SiV center. The 5-μm thickness has been chosen such that the development of the grain size can be followed from the interfacial layer on the substrate all the way to the top surface, where the grain size is expected to be larger than the confocal volume used to address SiV centers. The main goal is to study the correlation between the fluorescence background with the polycrystalline structure and the evolution of the grain size, which we address by confocal mapping of the Raman signal and the SiV fluorescence emission, and by scanning transmission electron microscopy (STEM). Moreover, we perform time-of-flight secondary ion mass spectrometry (ToF-SIMS) to investigate the chemical composition of the PCD membranes, with special attention to Si distribution and contamination.

## 2. Sample fabrication

We grow 5-μm thick diamond films by MPCVD using a 6-kW ellipsoidal plasma reactor as described in (36). Prior to deposition, the silicon substrate is seeded with diamond nanoparticles for high nucleation densities. It is then coated using a $CH_4/H_2$ gas mixture, at a substrate temperature of 800 °C, at 150 mbar gas pressure, and 300 sccm total gas flow rate. The concentration of $CH_4$ is 1 %. These parameters facilitate the growth of high-quality diamond films. After deposition, the diamond film is polished using conventional diamond polishing techniques to a surface roughness below 5 nm (rms). Then, circular grooves and windows are etched into the diamond film using deep reactive ion etching, resulting in self-standing PCD membranes with an unsupported area of 2 mm in diameter (see Fig. 1a). Finally, single PCD membranes are obtained by laser cutting along the circular grooves. The resulting samples consist of single polished PCD membranes of about 5-μm thickness stabilized by a silicon frame (see Fig. 1b).

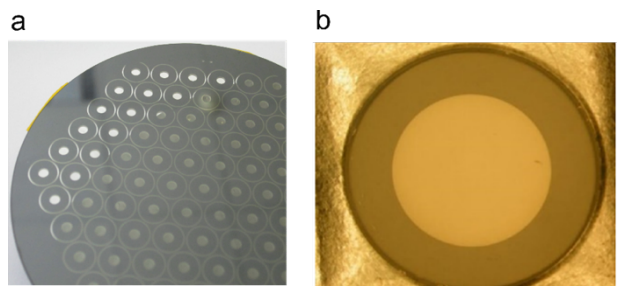

Figure 1: PCD membranes. (a) The image represents a batch of 5 μm-thick PCD membranes grown on an 8-inches Si wafer. (b) Microscope image of a laser segregated single PCD membrane with a diameter of 2 mm.

The PCD membranes are implanted at our recently developed facility, which consists of a High Voltage Tandem accelerator (3 MeV) along with a Negative sputter ion source (HVEE 860). A pulsed ion beam can be generated by the electrostatic deflector facility DEFEL (Electrostatic DEFlector) at the INFN LABEC laboratory in Florence (37). We are thus able to implant $Si^{3+}$ ions accelerated in the range of 0.2 – 15 MeV (according to the accelerating voltage and the state of charge of the selected ions) with a broad range of fluences, from $10^8$ to $10^{15}$ cm$^{-2}$. Moreover, using aluminum foils, we can control the implantation depth by degrading the ion energy down to a few tens of keV (35,38). To activate SiV color centers in the PCD membranes we perform an annealing step in an alumina oven at high vacuum conditions. The heating ramp starts from 20˚C to reach 1150˚C in 2 hours to maintain low pressure of the gases desorbed during heating, and the temperature stays at 1150˚C for 1hr.

## 3. Experimental results and discussion

Four different samples (from the same batch) are studied in order to identify the properties of the PCD membranes and the modifications introduced by ion implantation and annealing: PCD1 refers to a virgin sample; PCD2 refers to an annealed sample; PCD3 is an implanted and annealed sample (400 keV energy with a Si-ion fluence of $10^{-12}$ cm$^{-2}$, corresponding to a SiV depth distribution of 200-350 nm); PCD4 is an implanted and annealed sample (9 MeV energy with a Si-ion fluence of $10^{-8}$ cm$^{-2}$, due to the presence of two aluminum foils only 20% of the ions are implanted with a tail corresponding to a SiV depth distribution of 0-250 nm). Before the optical characterization, the PCD membranes are rinsed in a mixture of acetone and isopropanol (1:1) for 20 min. The diamond surface is also cleaned in a UV-ozone cleaner for 15 min to remove surface contaminations.

### 3.1 Optical setup and measurement approach

The optical properties of PCD membranes are explored using a home-built scanning confocal microscope (SCM) (see Fig. 2a). For the Raman and photoluminescence (PL) measurements, the sample is excited by a 656.15 nm laser diode (Pico Quant, LDH D-C-660 driven by PDL800-D), unless otherwise stated, using a high numerical-aperture (NA) objective (Olympus, MPlan Apo N 100x/0.95 NA). The same objective is used for both excitation and collection. The Raman and PL signals are carefully filtered by suppressing unwanted contributions (e.g., scattering, laser emission) using a confocal system. The sample is imaged using an EM-CCD camera (Princeton Instruments, ProEM-HS:512BX3). The spectrum is monitored using a spectrometer (Andor, Shamrock SR-500i-D2-SIL) equipped with an EM-CCD detector (Andor, Newton DU970P-BVF). Alternatively, the Raman and PL signals are detected by two avalanche photodiodes (APDs) (Micro Photon Devices, MPD-100-CTC) connected to time-correlated single-photon counting electronics (TCSPC) (Pico Quant, PicoHarp 300). To generate confocal maps, the sample is scanned in the *xy*-direction using a piezo stage (Physik Instrumente, P-542.2CL) following the scheme shown in Fig. 2b. To investigate the grain size evolution in the PCD membranes, mapping along the *z*-axis can be performed using a piezo objective scanner (Physik Instrumente, P-725 PIFOC) as shown in Fig. 2c.

The SCM mapping has an axial resolution of 1.4 µm and a lateral resolution of 0.4 µm, which allow scanning the PCD membrane at different depths and analyzing the grain size. The theoretical calculations for the axial and lateral resolution are made by assuming the SiV color centers in air. However, the presence of diamond (refractive index 2.4) could lead to optical aberrations and reduction of the axial resolution. To distinguish between the Raman and the PL signals, each APD is preceded by two different bandpass filters that generate two maps. Filter 1 is centered at 720 nm with 10 nm bandwidth (corresponding to the diamond Raman peak at 1332 cm$^{-1}$ (719 nm) (39)) and Filter 2 is centered at 740 nm with 13 nm bandwidth (corresponding to the ZPL of the SiV center at 738 nm (40)).

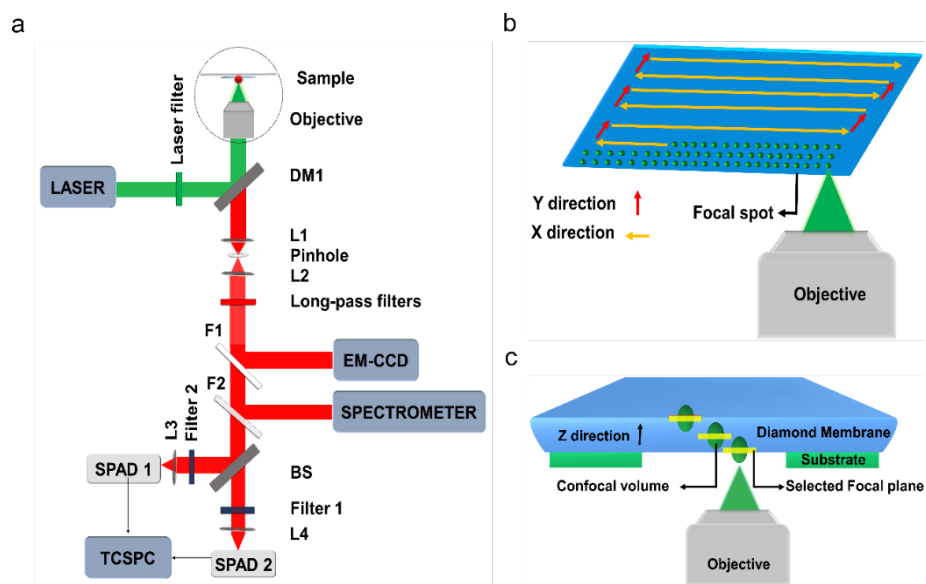

Figure 2: Optical setup and confocal scanning. (a) Scheme of the experimental setup for imaging, Raman, and PL analysis. The laser beam is expanded using a telescopic system. A dichroic mirror (DM1) directs the laser towards an objective to excite the sample. The same

objective collects the emission from the PCD membrane. DM1 transmits the Raman and PL signals from the sample and reflects the excitation laser. The signal is then filtered by a confocal system with two lenses ($L_1$, $L_2$), and a pinhole ($P_1$) (pinhole size 150 μm) and by a long-pass filter. A flipping mirror ($F_1$) in the optical path sends the light towards the EM-CCD camera for fluorescence imaging. The signal can also be sent towards the spectrometer using a flipping mirror ($F_2$). A 50/50 non-polarizing beam splitter (BS) sends the emitted photons to two APDs (SPAD 1 and SPAD 2) after focusing with the lenses ($L_3$ and $L_4$) for mapping. Two bandpass filters corresponding to the diamond Raman line (Filter 1) and the SiV ZPL (Filter 2) are placed in front of the APDs. The signal from the APDs is recorded using a TCSPC device. (b) *xy* mapping by scanning the sample over the objective. (c) *z* mapping by moving the objective.

Since the APDs do not allow us to distinguish the Raman and the SiV fluorescence signals from the background, we perform spectral measurements at certain regions on the sample surface to acquire the full spectral information. To identify the contributions of different spectral components to the overall emission, the signal at each point is analyzed at a 0.5 nm spectral resolution (150 l/mm grating, 357 nm bandpass). The intensity at Filters 1 and 2 represents the counts per second (cps) recorded by the APDs, which is then traced by integrating the recorded spectral signal over the wavelength range corresponding to the filters used in the mapping. Before performing this step, the background is identified by introducing a baseline corresponding to the counts averaged at the edges of the Filter 2 spectral window. Since the strongly inhomogeneous background prevents a proper fit of the SiV peak in some samples, this robust method allows an alternative estimation of the background. Finally, the background and the SiV signals are separately integrated, where the contribution below the baseline corresponds to the background fluorescence, and the one above the baseline indicates the SiV luminescence.

*3.2 Grain-size studies on PCD membranes*

The grain size-dependent physical and chemical properties of thick PCD films have already been studied (41). This section reports a comprehensive study of the optical properties of few-microns-thick PCD membranes via SCM maps generated from different depths. The diffraction-limited excitation laser that performs the SCM mapping on the PCD membrane starts from the top surface and it eventually approaches the surface close to the substrate (see Fig. 2c) using the piezo movement. Initially, the piezo performs a scan in the *z*-direction, which gives the information about the focal points corresponding to two surfaces of the PCD membrane in terms of count rates. When the confocal volume probes most of the diamond sample (surface close to the substrate), the count rate increases. Then, the home-built confocal set-up selects different planes for the depth-related scanning. The SCM maps from Filter 1 and Filter 2, shown in Figs. 3a and 3b, are generated from the top surface of PCD3, whereas Figs. 3c and 3d display the corresponding SCM maps obtained from the surface close to the substrate. The grain size decreases from about 3-5 µm near the top surface to less than 1 µm (diffraction limited) near the substrate. Moreover, the SCM maps generated from both filters show higher counts near the substrate. From the larger overlap of the confocal volume with the PCD membrane near the substrate, we can infer that Si contamination incorporated near the nucleation layer during the PCD growth creates additional SiV centers.

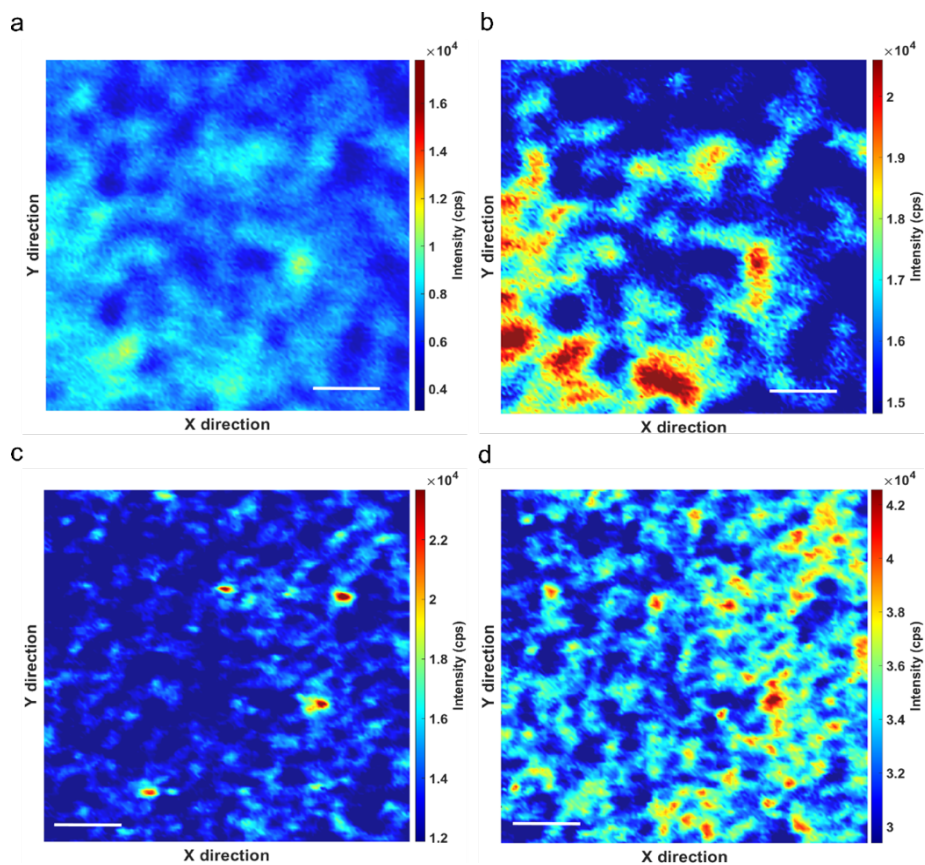

Figure 3: Optical characterization of the grain size in PCD membranes (PCD3). (a) and (b) SCM maps from Filter 1 and Filter 2, respectively, acquired on the top surface of the membrane. (c) and (d) SCM maps from Filter 1 and Filter 2, respectively, acquired near the substrate. The scale bar in the SCM maps is 5 µm and the step size is 0.2 µm.

To verify the grain size distribution estimated via the SCM maps at various depths, we performed STEM measurements (FEI Talos F200X, image acquisition at 200 kV) on the PCD1 membrane (the PCD membranes belong to the same batch). The TEM lamella is prepared by conventional focused ion beam (FEI Helios Nanolab 600) lift-out. Additionally, to a platinum (Pt) protective layer on top, a thin protective Au layer of only a few nm is applied at the bottom of the PCD membrane to ensure that the original bottom surface is preserved during milling. The bright-field STEM image shown in Fig. 4a points out that above the nucleation layer the crystal size grows rapidly, such that one can find diamond grains of nearly single crystalline quality with lateral dimensions of about 2 µm. The crystals are clearly visible due to the resulting Bragg contrast. The image shown in Fig. 4b obtained from one section of the PCD1 membrane confirms the growth of larger crystals immediately after the formation of an initial nucleation layer. The grain selection for the growth of larger grains takes place already after about 200 nm. Moreover, stacking faults and dislocations are rarely visible in this figure. The small discrepancy between the grain size inferred from the SCM maps and the STEM analysis must be attributed to the resolution of the confocal volume, which smears the grain boundaries and effectively increases the grain dimensions.

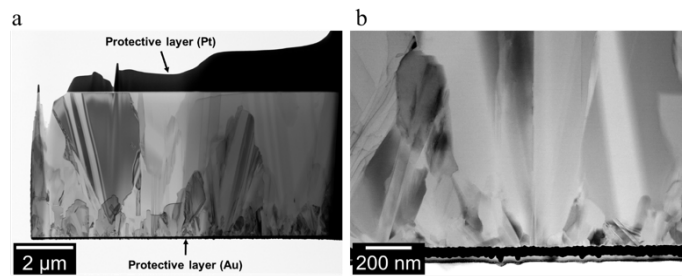

Figure 4: STEM characterization of the grain size in PCD membranes (PCD1). (a) Bright-field STEM image showing crystal growth, where the grain size is larger at the top surface (~ 2 µm grain size). (b) STEM image at the bottom of the diamond layer showing that grain selection takes place rapidly.

*3.3 Optical studies of PCD membranes*

*3.3.1 Virgin PCD membranes (PCD1)*

Figures 5a and 5b depict the SCM maps at Filter 1 and Filter 2 acquired near the upper surface of the PCD1 membrane, respectively. The grain size is about 3-5 µm; nonetheless, the GBs are still clearly visible in the SCM map of PCD1. The sample is not implanted, which means spectral features corresponding to the emission of SiV color centers are not expected unless there are diffused Si ions created during the growth process.

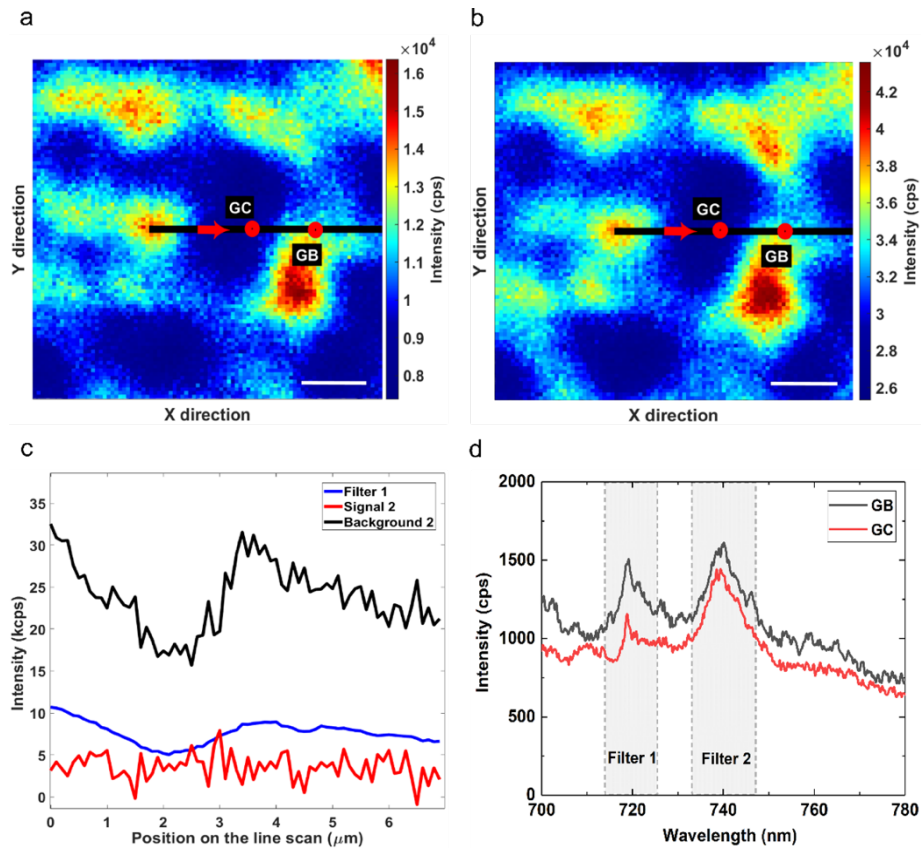

Figure 5: Optical characterization of a virgin PCD membrane (PCD1). (a) SCM map of the signal from Filter 1. The dark regions correspond to the GCs and the bright ones to the GBs. (b) SCM map of the signal from Filter 2. The scale bar in the SCM maps is 2 μm and the step size is 0.1 μm. (c) Spectral scanning performed at different positions (0.1 μm step size) along the black line indicated in Figs. 5a and 5b, showing the background around the ZPL emission of SiV centers (black curve), the background-subtracted PL from Filter 2 (red curve), and the signal from Filter 1 (blue curve). (d) PL spectra in the GC (red curve) and in the GB (black curve) taken at the corresponding red points marked in Figs. 5a and 5b. The bandpass of Filter 1 and Filter 2 is indicated.

The curves shown in Fig. 5c, obtained across the black line in Figs. 5a and 5b by processing the PL spectral as discussed in Sec. 3.1, indicate that the background-subtracted PL signal from Filter 2 (red curve) remains nearly constant, whereas the background PL signal from Filter 2 (black curve) and the PL signal from Filter 1 vary. They are maximal over the GB and drop near the GC. We thus consider the PL spectra at the GB (black curve) and at the GC (red curve) in Figure 5d. PCD1 exhibits a broad spectral feature at around 740 nm, which could be the signature of the SiV center, the GR1 center (42) or a combination of the two. Since this sample is not thermally treated nor ion-implanted, this contribution to the PL signal must be attributed to diffused Si ions introduced via the non-optimal growth conditions and to interstitial defects. In fact, the feature appears more prominent and broader at the GB. Lifetime measurements (not shown) confirm that the features near 740 nm can be associated to the SiV center, since the SiV and the GR1 centers have clearly different excited-state lifetimes (35,43). Additionally, the lower wavelength range presents three extra peaks associated with Raman scattering. The peak located at 710 nm corresponds to a Raman shift of 1150 cm$^{-1}$ and the peak at 719 nm is related to a Raman shift of 1332 cm$^{-1}$. The latter is the fingerprint of diamond (44–46) and it is clearly visible in both GC and GB. Moreover, the peak related to the 1150 cm$^{-1}$ shift is accompanied by a peak at 725 nm corresponding to a Raman shift of 1447.31 cm$^{-1}$. The origin of both 1150 cm$^{-1}$ and 1447.31 cm$^{-1}$ lines has been under debate for a long time (47). A. C. Ferrari et al. suggested that the peak at 1150 cm$^{-1}$ and the accompanying peak at 1447.31 cm$^{-1}$ are associated with trans-polyacetylene lying in GBs and not with amorphous/nanocrystalline diamond (48). It is clear from Fig. 5d that in the GC the band at 1447.31 cm$^{-1}$ (725 nm) is barely visible and that the PL background is also lower, indicating that this region is mostly free from impurity carbon phases (sp$^2$ carbon phases). On the other hand, in the GB, the signature Raman line at 1332 cm$^{-1}$ (719 nm) has emerged with a 1447.31 cm$^{-1}$ (725 nm) line. Moreover, a Raman shift of 1560.53 cm$^{-1}$ (731 nm), which is the G mode of graphite (49), is more present in the GBs due to the existence of sp$^2$ carbon. Hence, the sp$^3$ phase of carbon content appears higher in the GC than in the GB. We would like to remark that a fraction of these Raman lines is included in Filter 1. Therefore, the SCM map associated with Filter 1 is not only representing the intensity of the diamond Raman band, as originally intended. That is why we combine the SCM mapping with a spectral analysis and we do so also in the following samples.

*3.3.2 Un-implanted, thermally annealed PCD membranes (PCD2)*

The SCM maps from Filter 1 (Fig. 6a) and Filter 2 (Fig. 6b) display the GCs and the GBs near the upper surface of the PCD2 membrane. Figure 6c reports the spectral measurements performed along the black lines marked in Figs. 6a and 6b, which have been processed as discusses in Sec. 3.1 to monitor the variations of the PL background signal in the spectral window of Filter 2 (black curve), the PL intensity from Filter 1 (blue curve), and the background-subtracted PL intensity from Filter 2 (red curve). A significant remark is that both PL background and signal from Filter 2 are larger than the PL at Filter 1. On the one hand, thermal activation suppresses the spectral peaks associated with trans-polyacetylene, although a small signature remains in the GB, as shown by the PL spectra in Fig. 6d (acquisition points corresponding to GC and GB marked in Figs. 6a and 6b). Moreover, the presence of the *G* band is particularly strong at the GB. The strong spectral feature at 738 nm suggests the formation of SiV centers via diffused Si ions in the sample, both at the GC and at the GB. Repeated measurements performed on the GBs and GCs reproduced the results shown in Figs. 6c and 6d.

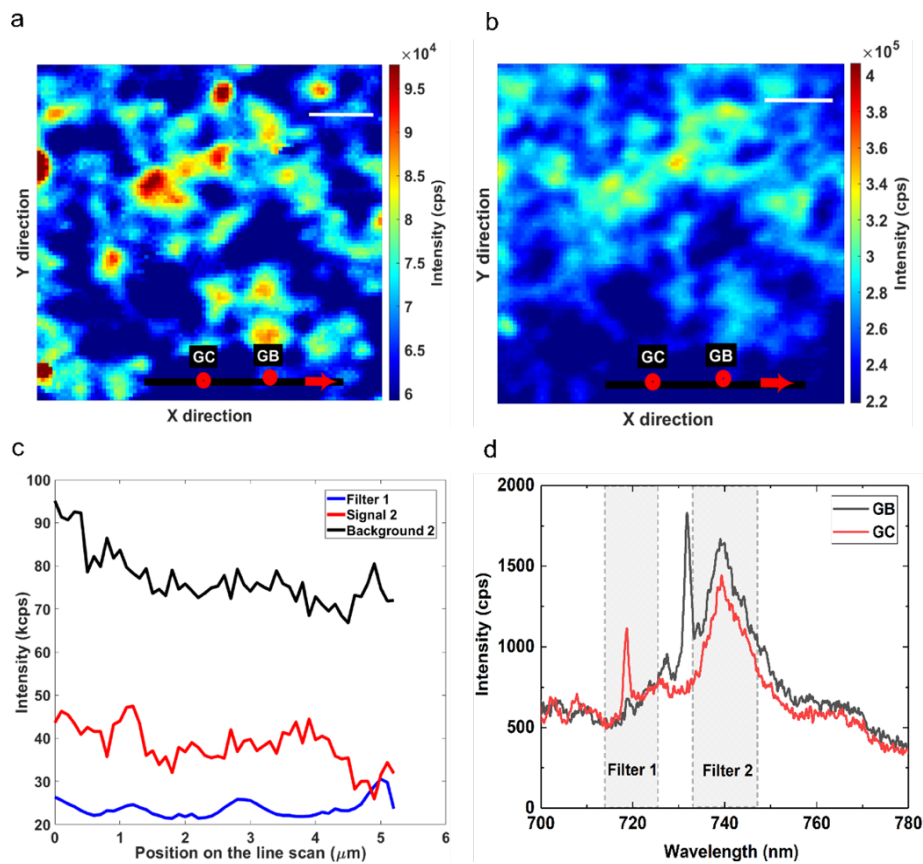

Figure 6: Optical characterization of un-implanted and annealed PCD membranes (PCD2). (a) SCM map of the signal from Filter 1. The dark regions correspond to the GCs and the bright ones to the GBs. (b) SCM map of the signal from Filter 2. The scale bar in the SCM maps is 2 µm and the step size is 0.1 µm. (c) Spectral scanning performed at different positions (0.1 µm step size) along the black line indicated in Figs. 6a and 6b, showing the background around the ZPL emission of SiV centers (black curve), the background-subtracted PL from Filter 2 (red curve) and the signal from Filter 1 (blue curve). (d) PL spectra in the GC (red curve) and in the GB (black curve) taken at the corresponding red points marked in Figs. 6a and 6b. The bandpass of Filter 1 and Filter 2 is indicated.

*3.3.3 Implanted and annealed PCD membranes (PCD3)*

Finally, we study the optical properties of implanted PCD membranes (PCD3). Figure 7a is the SCM map near the upper surface of PCD3 corresponding to the intensity from Filter 1. The SCM map from Filter 2 is shown in Fig. 7b. The fact that the SCM map does not exhibit a substantially larger signal than the one in Fig. 6b indicates that overall the SiV centers created by ion implantation are not overwhelming the PL signal. In the curves of Fig. 7c, which report spectral measurements processed as discussed in Sec. 3.1, both background and PL signals from Filter 2 (black and red curves, respectively) increase at the GB and drop at the GC, as observed for the previous two situations. Moreover, the integrated counts at Filter 1 (blue curve) are less compared to those at Filter 2, because of the SiV center activation. Different iterations of the spectral line scan on the PCD3 sample showed similar results verifying the findings. The PL spectrum of SiV centers at the GB and at the GC (marked as red circles in Figs. 7a and 7b) is presented in Fig. 7d. The PL background is enhanced due to the presence of impurities formed during diamond deposition followed by ion-implantation and thermal annealing although, compared to GCs, it is larger at the GBs. Moreover, the background appears more pronounced than in PCD2, probably due to ion implantation. Comparing to the PCD1, the post-annealing process performed on PCD3 has quenched the spectral peaks represented by trans-polyacetylene (at 710 nm and at 725 nm) lying in GBs and GC. Because of the higher $sp^2$ phase impurities accumulated at the GB, the *G* band at 731 nm correlated with the disordered graphitic carbon is noticeable in the spectra at the GBs, while it is completely suppressed in the GC.

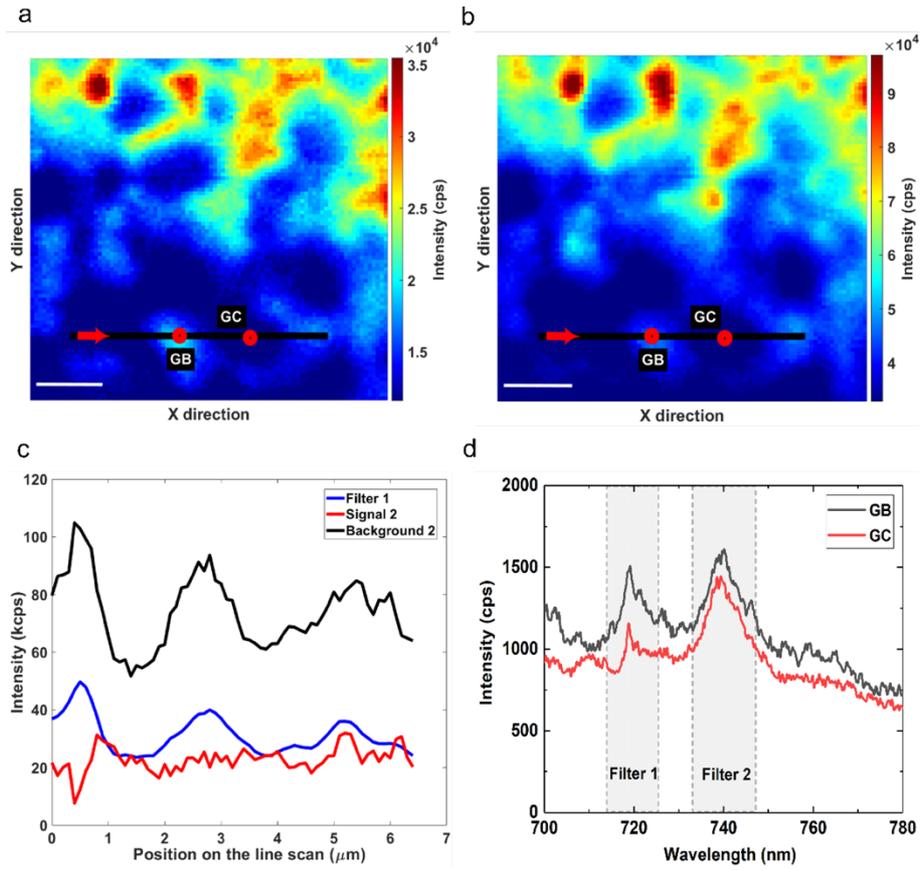

Figure 7: Optical characterization of implanted and annealed PCD membranes (PCD3). (a) SCM map of the signal from Filter 1. The dark regions correspond to the GCs and the bright ones to the GBs. (b) SCM map of the signal from Filter 2. The scale bar in the SCM maps is 2 μm and the step size is 0.1 μm. (c) Spectral scanning performed at different positions (0.1 μm step size) along the black line indicated in Figs. 5a and 5b, showing the background around the ZPL emission of SiV centers (black curve), the background-subtracted PL from Filter 2 (red curve), and the signal from Filter 1 (blue curve). (d) PL spectra in the GC (red curve) and in the GB (black curve) taken at the corresponding red points marked in Figs. 5a and 5b. The bandpass of Filter 1 and Filter 2 is indicated.

In order to further understand the origin of the increased background and optimize the PL signal of the SiV color centers, we investigate the PCD1 and PCD3 membranes using 532 nm and 690 nm laser excitation (not shown). The nitrogen-vacancy (NV) related background is evident in the spectra of both samples, when excited by the 532 nm laser. These features are strongly suppressed, when the samples are excited by the 690 nm laser, as NV centers have significantly less absorption in this spectral range. However, the PCD3 membrane exhibits a higher background as compared to PCD1, as shown by Figs. 5 and 7. This can be ascribed to the creation of NV centers and other defects, since ion implantation creates vacancies and thermal annealing favors the formation of color centers (SiV, but also NV, if nitrogen is present in the diamond sample). This suggests that effectively reducing the nitrogen concentration during MPCVD growth can improve the signal-to-noise ratio required for the detection of single SiV color centers in PCD membranes.

*3.4 Determination of Si distribution in un-implanted and implanted PCD membranes (PCD1 and PCD4) by Time-of-Flight Secondary Ion Mass Spectrometry*

To determine the Si distribution through PCD membranes we perform ToF-SIMS depth profiling on both un-implanted (PCD1) and implanted (PCD4) samples. The technique is suitable for investigation of low concentrated species with a high depth resolution (50). Depth profiles are recorded on a ToF-SIMS V (Ion. ToF, Münster, Germany) using a 25 keV $Bi^+$ ion beam bunched down to <0.8 ns in negative polarity. Two sputter regimes are used: high rate for investigation of the entire membrane (Cs 2kV, 150 x 150 µm crater size with 30 x 30 µm measurement spot in the center of the crater) and low rate for a comparison of implantation depth (Cs 2kV, 250 x 250 µm crater size with 50 x 50 µm measurement spot in the center of the crater). At the top surface, PDMS containing contaminations are observed for all samples (not shown), probably originating from sample handling and transport. In negative ToF-SIMS spectra, the characteristic signals for diamond $C_x^-$; $C_9^-$, and $C_{10}^-$ are chosen for display as their intensities are comparable to other signals of interest. The composition of the PCD membranes is observed to be homogeneous over the entire thickness (see Fig 8a, normalized to $C_{10}^-$ to account for fluctuations of the ion beam), with the exception of both interfaces. Within the un-implanted sample (PCD1), no Si-containing signals are detected, which indicates no significant contamination with Si in the PCD membrane prior to implantation. The bottom interface of the sample is examined to investigate diffusion of Si from the removed substrate, caused by the growth process of diamond, see Fig. 8b. The profile is cut off at the signal level of spectral noise and not normalized (at interfaces, normalization induces artefacts). The appearance of $SiO_2^-$ and $Si^-$ signals after the decay of $C_x^-$ signals indicates a non-complete removal of the substrate. The overlap area of these signals can be explained by either surface roughening induced during the sputter depth profiling process, diffusion of substrate into the diamond or a combination of both effects. Thus, a maximum diffusion depth of less than 100 nm is determined for Si migrating into the PCD1 membrane during the growth process (shaded area in Fig. 8b). In combination with the surface roughening induced by the substrate removal, as observed in Fig. 4b, this indicates that no significant amount of Si is incorporated from the silicon substrate.

The presence of Si at the top surface is compared for both implanted (fluence $10^8$ cm$^{-1}$) and un-implanted PCD membranes (PCD1 and PCD4). In Fig. 8c the profiles are normalized to the $C_{10}^-$ signals for quantitative comparison. The $C_9^-$ signals overlap perfectly, indicating that the profiles are comparable. At the surface (~ 20 nm), the Si$^-$ signal is caused by the presence of PDMS contamination. The tail in the vanishing signal of Si$^-$ on the un-implanted sample (PCD1) stems from a non-homogeneous thickness of the PDMS contamination and sputter-induced surface roughening. In the depth profile the Si$^-$ signal is clearly enhanced on the implanted sample (PCD4) for ~ 200-250 nm sputter depth, which is consistent with the implantation conditions (tail with a penetration up to 250 nm estimated by SRIM simulations (51)).

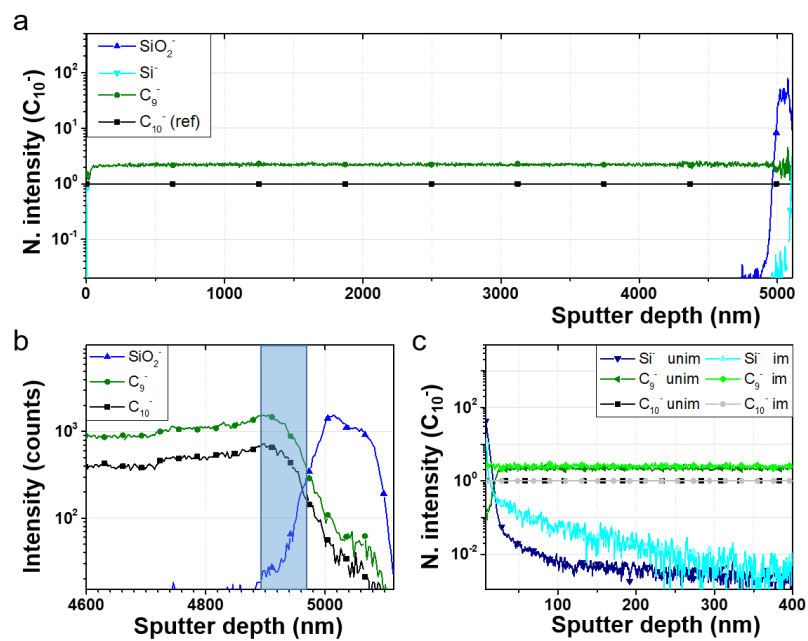

Figure 8: ToF-SIMS analysis of un-implanted and implanted PCD membranes (PCD1 and PCD4), depth profiles in negative polarity. (a) Entire PCD1 depth profile, where the marked area shows the maximum diffusion depth of the substrate, (b) magnification of the bottom interface of PCD1 (signal intensity not normalized) and (c) overlay of the top interfaces of PCD1 (unim) and PCD4 (im).

These findings indicate that by careful etching of the top and bottom interfaces, it is possible to eliminate any remaining Si contamination in PCD membranes, so that the creation of SiV centers can be obtained in a controlled manner by Si-ion implantation.

## 4. Conclusions

We investigate the properties of PCD membranes containing SiV centers to identify procedures, which eventually would allow addressing single emitters. We do so by performing a compelling study that combines microphotoluminescence measurements, electron microscopy and elemental analysis with depth information on the same batch of samples, hence providing an accurate correlation between the results obtained by the different techniques. Firstly, the dimension of the diamond grains is crucial for the interrogation of single SiV centers away from the background-rich GBs. To this end, we probe the evolution of the grain size by confocal scanning and STEM analysis, showing that near the top surface the dimensions of the diamond crystalline grains would be sufficiently large to address SiV centers by confocal microscopy, without undesired excitation at the GBs (according to the achievable resolution set by diffraction). Next, we correlate the Raman signal with the SiV emission and the luminescence background for different types of PCD membranes (unprocessed, annealed, implanted and annealed). Because the signal at Filter 1 contains several Raman peaks, a spectral analysis is necessary to identify the Raman signal from diamond. While annealing reduces the background but activates unwanted SiV centers originating from the Si contamination, the implantation process tends to increase the background due to the creation of vacancies that combine with Si and nitrogen impurities in the subsequent annealing step. Finally, we perform a ToF-SIMS analysis of un-implanted and implanted PCD membranes that clarify the depth profile of the Si distribution and contamination. In practice, the latter is closely confined to the interfaces and thus it can be removed by controlled etching. Our study indicates that PCD membranes with a thickness of a few microns could be suitable for quantum optical applications, provided that the nano-crystalline regions are removed, for instance by back thinning, or if the spatial resolution in the optical excitation is substantially increased, for instance by near-field or antenna-enhanced microscopy (52).

**Acknowledgments**

The authors gratefully acknowledge financial support from the University of Siegen and the German Research Foundation (DFG) (INST 221/ 118-1 FUGG, 410405168). The authors also acknowledge INFN-CHNet, the network of laboratories of the INFN for cultural heritage, for support and precious contributions in terms of instrumentation and personnel. S. Lagomarsino, N. Gelli, S. Sciortino, and L. Giuntini wish to thank F. Taccetti for experimental assistance and suggestions. H. Kambalathmana acknowledges support from P. Reuschel and N. Soltani. M.S. Killian would like to thank C. Hasenest for depth calibration assistance. M. Agio would like to thank C. Becher and H. Galal for helpful discussions. This work is based upon networking from the COST Action MP 1403 "Nanoscale Quantum Optics," supported by COST (European Cooperation in Science and Technology). Part of this work was performed at the Micro- and Nanoanalytics Facility (MNaF) of the University of Siegen.